\documentclass[prb,twocolumn,showpacs,amsmath,amssymb,superscriptaddress]{revtex4}
\usepackage{graphicx}
\usepackage{bm}

\begin{document}

\title{Dynamic behaviour of Josephson-junction qubits: crossover between
Rabi oscillations and Landau-Zener transitions}

\author{S.N. Shevchenko$^{(1,2)}$, A.S. Kiyko$^{(1)}$, A.N. Omelyanchouk$%
^{(1)}$, W. Krech$^{(2)}$ \\
$^{(1)}$ B. Verkin Institute for Low Temperature Physics and Engineering, \\
47 Lenin Ave., 61103, Kharkov, Ukraine.\\
$^{(2)}$ Friedrich Schiller University, Institute for Solid State Physics, \\
Helmholtzweg 5, D-07743 Jena, Germany.}

\date{\today}

\begin{abstract}
We study the dynamic behaviour of a quantum two-level system with periodically varying
parameters by solving the master equation for the density matrix. Two limiting cases
are considered: multiphoton Rabi oscillations and Landau-Zener (LZ) transitions. The
approach is applied to the description of the dynamics of superconducting qubits. In
particular, the case of the interferometer-type charge qubit with periodically varying
parameters (gate voltage or magnetic flux) is investigated. The time-averaged energy
level populations are calculated as functions of the qubit's control parameters.
\end{abstract}

\pacs{03.67.Lx, 03.75.Lm, 74.50.+r, 85.25.Am}

\maketitle

\section{Introduction}

The quantum two-level system is a model to describe a number of physical objects such
as atoms, quantum dots, molecular magnets, etc. The system can be excited from the
ground state to the upper state by changing in time its parameters (due to external
fields). If the parameters vary adiabatically
slow, the excitation mechanism is called Landau-Zener (LZ) transition \cite%
{LZS}$^{,}$ \cite{BVK}; if the amplitude of the field is small and its frequency
is comparable to the level distance, Rabi oscillations occur \cite%
{Rabi}$^{,}$ \cite{DelKrai}. Thus the occupation of the levels in a two-state system
can be controlled by several parameters, e.g., by amplitude and frequency of an
external field \cite{GrifHang}.

Recently, the two-level model was used to describe Josephson-junction systems
\cite{MShSh}, i.e., both charge qubits \cite{chargeQbit} and flux qubits
\cite{fluxQbit}. The subject of the present work is the investigation of the dynamic
behaviour of superconducting qubits with periodically swept parameters, which is
important from the point of view of state control and readout. We will relate our
results to some other articles
concerning the resonant excitation of a two-level system \cite%
{KrainovYakovlev}$^{-}$\cite{TerNak} which was shown to be relevant for superconducting
qubits too \cite{Nakamura01}$^{-}$\cite{Born}. Particularly, we will describe the
dynamic behaviour of the interferometer-type charge qubit
\cite{Falci}$^{-}$\cite{Krech02}.

To study theoretically the dynamic behaviour of a Josephson qubit, we
make use of the master equation for the density matrix rather than the Schr%
\"{o}dinger equation because this allows to take into account both relaxation and
non-zero temperature effects. (However, in this paper we will assume the
zero-temperature limit, because we are interested in the influence of relaxation
processes only on the qubit dynamics.)

The diagonal components of the density matrix define the probabilities of finding the
system in the respective states of the basis in which the density matrix is presented.
Thus, executing the calculations we ought to deal with a particular basis. Our
calculations are mostly carried out in the stationary basis $\{\left\vert
-\right\rangle ,\left\vert +\right\rangle \}$ of the eigenstates of the Hamiltonian
$\widehat{H}^{(0)}$\ in the absence of time-dependent terms. We do that for two
reasons. First, it is convenient to describe Rabi oscillations and multiphoton
transitions. Second, in the case of the charge qubit these states $\{\left\vert
-\right\rangle ,\left\vert +\right\rangle \}$ are eigenstates of the current operator,
which is related to the experimentally measurable values \cite{Krech02}. The important
point is that we can get the occupation probability of any state provided we know the
probabilities for the states in a particular basis. To demonstrate this, we will change
over from the stationary basis to the so-called adiabatic basis (consisting of the
instantaneous eigenstates of the time-dependent Hamiltonian $\widehat{H}$) to describe
the LZ effect. We emphasize that the presented results are valid for the description of
any two-level system with periodically swept parameters, particularly, of a
superconducting qubit. In a Josephson-junction qubit the gate voltage or the magnetic
flux can be modulated periodically.

The paper is organized as following. In Sec. II the basic equations are presented. In
Sec. III we study multiphoton processes and LZ transitions in a two-level system with
time-dependent parameters. We apply the respective results to the phase-biased charge
qubit in Sec. IV.

\section{The basic equations}

We start from the Hamiltonian of a two-level system (see e.g. Ref. 6)%
\begin{equation}
\widehat{H}^{(0)}=-\frac{B_{x}^{(0)}}{2}\widehat{\sigma }_{x}-\frac{%
B_{z}^{(0)}}{2}\widehat{\sigma }_{z}
\end{equation}%
in the basis of "physical"\ states $\{\left\vert 0\right\rangle ,\left\vert
1\right\rangle \}$, where $\widehat{\sigma }_{z}\left\vert 0\right\rangle =\left\vert
0\right\rangle $, $\widehat{\sigma }_{z}\left\vert 1\right\rangle =-\left\vert
1\right\rangle $, $\widehat{\sigma }_{x,y,z}$ are the Pauli matrices. The "physical"\
states are the
eigenstates of the Hamiltonian $\widehat{H}^{(0)}$ for $%
B_{x}^{(0)}/B_{z}^{(0)}\rightarrow 0$. In the case of a charge qubit, these states
correspond to a definite number of Cooper pairs on the island. For a flux qubit, they
correspond to a definite direction of the current circulating in the ring. The
Hamiltonian (1) is diagonalized by means of the matrix
\begin{equation*}
\widehat{S}=\exp \left( i\frac{\lambda }{2}\widehat{\sigma}_{y}\right) =\left[
\begin{array}{cc}
\cos \lambda /2 & \sin \lambda /2 \\
-\sin \lambda /2 & \cos \lambda /2%
\end{array}%
\right] ,
\end{equation*}%
where
\begin{equation*}
\sin \lambda =-\frac{B_{x}^{(0)}}{\Delta E},\quad \cos \lambda =\frac{%
B_{z}^{(0)}}{\Delta E},\quad \Delta E=\sqrt{B_{x}^{(0)2}+B_{z}^{(0)2}}.
\end{equation*}%
The eigenstates $\left\vert -\right\rangle \ $and $\left\vert +\right\rangle $ of the
time-independent Hamiltonian $\widehat{H}^{(0)}$ are connected with the initial basis:
\begin{equation*}
\left[
\begin{array}{c}
\left\vert -\right\rangle  \\
\left\vert +\right\rangle
\end{array}%
\right] =\widehat{S}\left[
\begin{array}{c}
\left\vert 0\right\rangle  \\
\left\vert 1\right\rangle
\end{array}%
\right] \text{.}
\end{equation*}

Next, we introduce the time-dependent terms into the Hamiltonian (1),
\begin{equation*}
\widehat{H}^{(0)}\rightarrow \widehat{H}=\widehat{H}^{(0)}+\widehat{H}%
^{(1)}(t).
\end{equation*}

We consider two situations,
\begin{equation}
(a):\text{ \ \ \ }B_{x}=B_{x}^{(0)},\text{ \
}B_{z}=B_{z}(t)=B_{z}^{(0)}+B_{z}^{(1)}(t), \label{1}
\end{equation}%
\begin{equation}
(b):\text{ \ \ \ }B_{z}=B_{z}^{(0)},\text{ \
}B_{x}=B_{x}(t)=B_{x}^{(0)}+B_{x}^{(1)}(t), \label{2}
\end{equation}%
where the time-independent/dependent terms are marked with $(0)/(1)$
indices.\ Making use of the transformation $\widehat{H}^{/}=\widehat{S}^{-1}%
\widehat{H}\widehat{S}$, we get the Hamiltonian $\widehat{H}^{/}$ in the energy
representation $\{\left\vert -\right\rangle ,\left\vert +\right\rangle \}$
corresponding to these cases:
\begin{equation}
\widehat{H}_{a}^{/}=-\frac{\Delta E}{2}\widehat{\tau }_{z}-\frac{%
B_{z}^{(1)}(t)}{2}\left( \sin \lambda \widehat{\tau }_{x}+\cos \lambda \widehat{\tau
}_{z}\right) ,  \label{H1}
\end{equation}%
\begin{equation}
\widehat{H}_{b}^{/}=-\frac{\Delta E}{2}\widehat{\tau }_{z}-\frac{%
B_{x}^{(1)}(t)}{2}\left( \cos \lambda \widehat{\tau }_{x}-\sin \lambda \widehat{\tau
}_{z}\right) .  \label{H2}
\end{equation}%
For convenience, we use different notations for the Pauli matrices, $%
\widehat{\sigma }_{i}$\ and $\widehat{\tau }_{i}$, which operate in the
bases $\{\left\vert 0\right\rangle ,\left\vert 1\right\rangle \}$ and $%
\{\left\vert -\right\rangle ,\left\vert +\right\rangle \}$, respectively.

We emphasize that after the substitution
\begin{align}
B_{x}^{(1)}(t)& \rightarrow B_{z}^{(1)}(t),  \label{substitute} \\
B_{z}^{(0)}& \rightarrow -B_{x}^{(0)},  \notag \\
B_{x}^{(0)}& \rightarrow B_{z}^{(0)}  \notag
\end{align}%
the problem (b) coincides with problem (a). This transition from Eq. (\ref{H2}) to Eq. (\ref{H1}) corresponds to a $%
\pi /2$ rotation about the $y$-axis.

To unify expressions (\ref{H1}) and (\ref{H2}) to get the equations for numerical
calculations, we write down the Hamiltonian $\widehat{H}^{/}$ as following:
\begin{equation}
\widehat{H}^{/}=\frac{A}{2}\widehat{\tau }_{x}+\frac{C}{2}\widehat{\tau }%
_{z}.  \label{H}
\end{equation}

The quantum dynamics of our two-level system can be characterized within the standard
density-matrix approach \cite{Blum}. The time-evolution of the total system composed of
the two-level system and the reservoir is described by the Liouville equation. After
tracing over the reservoir variables, the Liouville equation can be simplified to the
so-called master equation for the reduced density matrix $\widehat{\rho }$. It can be
written in the form
\begin{equation*}
\widehat{\rho }=\frac{1}{2}\left[
\begin{array}{cc}
1+Z & X-iY \\
X+iY & 1-Z%
\end{array}%
\right]
\end{equation*}%
which ensures the condition $Tr\widehat{\rho }$ $=1$. The effect of relaxation
processes in the system due to the weak coupling to the reservoir
can be phenomenologically described with the dephasing rate $\Gamma _{\phi }$%
\ and the relaxation rate $\Gamma _{relax}$(\cite{foot1}). Then the master equation
takes the form of Bloch equations \cite{Blum}:
\begin{align}
\frac{dX}{dt}& =-CY-\Gamma _{\phi }X,  \label{eq1} \\
\frac{dY}{dt}& =-AZ+CX-\Gamma _{\phi }Y,  \label{eq2} \\
\frac{dZ}{dt}& =AY-\Gamma _{relax}\left( Z-Z(0)\right) .  \label{eq3}
\end{align}%
(Throughout the paper we admit $\hbar =1$.) From these equations we get $Z(t) $ which
defines the occupation probability of the upper level $\left\vert +\right\rangle $,
$P_{+}(t)=\rho _{22}(t)=\frac{1}{2}(1-Z(t))$. We choose the initial condition to be
$X(0)=Y(0)=0$, $Z(0)=1$, that corresponds to the system in the ground state $\left\vert
-\right\rangle $. We have calculated the time evolution of $P_{+}(t)$ (which is
essential e.g. for the snap-shot measurements \cite{Aasime}) as well as the
time-averaged probability $\overline{P}_{+}$ (which is essential e.g. for the impedance
measurement technique \cite{Ilichev04}). We note that the asymptotic expression,
$\left. P_{+}(t)\right\vert _{t\rightarrow \infty }$, is periodic in time with the
period $T=2\pi /\omega $ of the time-dependent term of the Hamiltonian
$\widehat{H}^{(1)}(t)$ (see Sec. 12 of Ref. 5).

\section{Non-stationary effects in a two-level system}

\subsection{Rabi oscillations}

Hereafter we will treat the\ problem (a) (see Eq. (\ref{1})), making use of the
following notations:
\begin{equation}
\widehat{H}=\Delta \widehat{\sigma }_{x}+x(t)\widehat{\sigma }_{z},\text{ \ }%
x(t)=x_{off}+x_{0}\sin \omega t  \label{Ham_ShIF}
\end{equation}%
with
\begin{align}
B_{x}^{(0)}& =-2\Delta , \\
B_{z}^{(0)}& =-2x_{off},  \notag \\
B_{z}^{(1)}(t)& =-2x_{0}\sin \omega t.  \notag
\end{align}%
This reformulation is convenient for comparing our results with the results of other
papers \cite{KrainovYakovlev}$^{-}$\cite{ShIF}. Then the Hamiltonian (\ref{H1}) can be
rewritten:
\begin{align}
\widehat{H}^{/}& =-\frac{\Delta E}{2}\widehat{\tau }_{z}+x_{0}\sin \omega
t\cdot \widehat{V},  \label{Ham_Rabi} \\
\widehat{V}& =\frac{2\Delta }{\Delta E}\widehat{\tau }_{x}-\frac{2x_{off}}{%
\Delta E}\widehat{\tau }_{z}.  \notag
\end{align}

\begin{figure}[t]

\includegraphics[width=8.5cm]{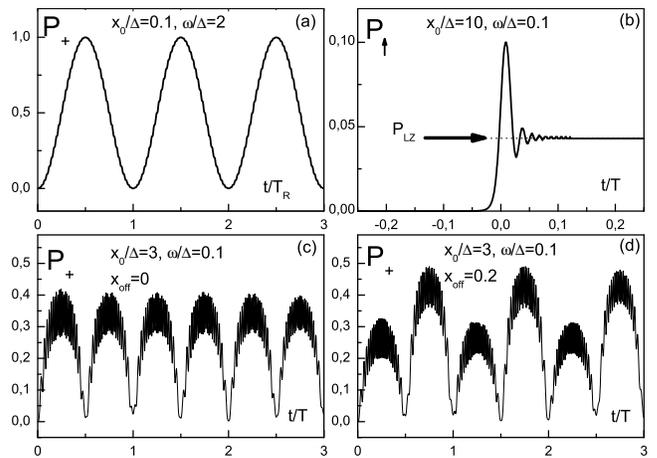}
\caption{Time dependence of upper-level occupation probabilities. (a) Rabi oscillations
in $P_{+}$ with the period $T_{R}=2\protect\pi /x_{0}$, (b) LZ transition in
$P_{\uparrow }$ (see Sec. III.B), (c) and (d) $P_{+}$ probability evolution in the case
of periodically
swept parameters at $x_{off}=0$ and $x_{off}\neq 0$. Here $\Gamma _{\protect%
\phi }=\Gamma _{relax}=0$; $T=2\protect\pi /\protect\omega $.} \label{P(t)}
\end{figure}

\begin{figure}[t]
\includegraphics[width=8.8cm]{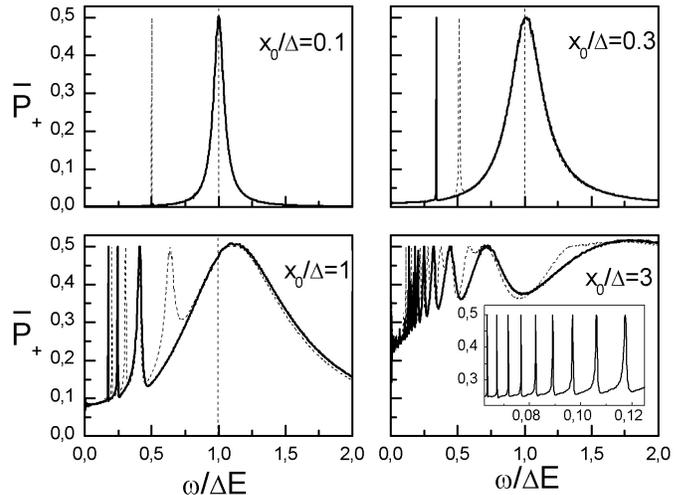}
\caption{Dependence of the probability $\overline{P}_{+}$ on the frequency
$\protect\omega $ for
different $x_{0}$ at $\Gamma _{\protect\phi }=\Gamma _{relax}=0$ and at $%
x_{off}=0$ (solid line) and $x_{off}=0.2\Delta $ (dashed line). (Only the first few
resonant peaks are plotted; the others, which are very narrow, are not shown at the
graphs.) Inset: enlargement of the low-frequency region.} \label{P(w)}
\end{figure}

First consider the situation $x_{off}=0$, then the difference between the stationary
energy levels is $\Delta E=2\Delta $. In the case
\begin{equation}
\Delta \omega \equiv \omega -\Delta E\ll \Delta E\text{ and }x_{0}\ll \Delta E,
\label{D_w}
\end{equation}%
one can use the so-called rotating wave\ approximation, and the result is
\cite{Flugge}%
\begin{equation}
P_{+}(t)=\frac{1}{2}\frac{x_{0}^{2}}{x_{0}^{2}+(\Delta \omega )^{2}}\left( 1-\cos
\sqrt{x_{0}^{2}+(\Delta \omega )^{2}}t\right) .  \label{P(t)_Rabi}
\end{equation}%
For the average probability, it follows
\begin{equation}
\overline{P}_{+}=\frac{1}{2}\frac{x_{0}^{2}}{x_{0}^{2}+(\Delta \omega )^{2}}.
\label{P_av_Rabi}
\end{equation}%
This means that at $\omega =\Delta E$ there is resonance, $\overline{P}_{+}=%
\frac{1}{2}$, and $P_{+}(t)$ is an oscillating function with the frequency $%
x_{0}$. This is illustrated in Fig. \ref{P(t)}a. The width of the peak at $%
\omega =\Delta E$ of the $\overline{P}_{+}$--$\omega $ curve at the half-maximum (i.e.,
at $P_{+}=1/4$) is approximately $2x_{0}$ (see the upper panels of Fig. \ref{P(w)}).

Resonant excitations of a two-level system for $x_{0}/\Delta E\ll 1$ may
occur not only at $\omega =\Delta E$. In the $K$-th order approximation, $%
(x_{0}/\Delta E)^{K}$, resonances exist at $\omega \simeq \Delta E/K$ \cite%
{DelKrai}$^{,}$ \cite{CohTan}. For $x_{off}=0,$ the resonances occur at odd $K$ only.
The dependence of $P_{+}$ on time in the vicinity of $K$-th resonance is described by a
relation similar to Eq. (\ref{P(t)_Rabi}) (with substitutions $\Delta \omega
\rightarrow \Delta \omega ^{(K)}\simeq K\omega
-\Delta E$ \ and $x_{0}\rightarrow $ $x_{0}^{K}$). Hence, the width of the $%
K $-th resonance is of the order of $x_{0}^{K}$. In the resonance, $\Delta
\omega ^{(K)}=0$, the energy levels $\left\vert -\right\rangle $ and $%
\left\vert +\right\rangle $ are equally populated so that $\overline{P}%
_{+}(\omega =\omega ^{(K)}\simeq \Delta E/K)=1/2$, which is the maximum value of
$\overline{P}_{+}$, since the population inversion in a two-level system is not
possible.

When $x_{off}\neq 0$, the matrix $\widehat{V}$ has non-zero diagonal elements (see Eq.
(\ref{Ham_Rabi})). The appearance of the non-zero diagonal elements does not influence
Rabi resonances at odd $K$, but rather results in the generation of the resonances at
even $K$. This is demonstrated in Fig. \ref{P(w)} and discussed in Sec. III.C.

\subsection{LZ effect}

The LZ effect manifests in the non-adiabatic transition with the probability
\begin{equation}
P_{LZ}=\exp \left( -\frac{\pi \Delta ^{2}}{\omega x_{0}}\right)  \label{P_LZ}
\end{equation}%
between two adiabatic energy levels during a single-sweep event \cite{LZS}.

The time-dependent Hamiltonian $\widehat{H}$ is diagonalized in the adiabatic basis,
denoted as $\{\left\vert \downarrow \right\rangle ,\left\vert \uparrow \right\rangle
\}$, by the $\widehat{S}_{1}$ matrix,
\begin{gather*}
\widehat{S}_{1}=\left[
\begin{array}{cc}
\cos \frac{\eta }{2} & \sin \frac{\eta }{2} \\
-\sin \frac{\eta }{2} & \cos \frac{\eta }{2}%
\end{array}%
\right] , \\
\sin \eta =-\frac{B_{x}}{\Delta E_{1}}=\frac{2\Delta }{\Delta E_{1}},\text{ }%
\cos \eta =\frac{B_{z}}{\Delta E_{1}}=-\frac{2x(t)}{\Delta E_{1}}, \\
\Delta E_{1}(t)=2\sqrt{\Delta ^{2}+x(t)^{2}}.
\end{gather*}%
The instantaneous eigenvalues of $\widehat{H}$ are $E_{\downarrow ,\uparrow
}=\pm \Delta E_{1}(t)/2$.

We now can change over from the stationary basis to the adiabatic one,
\begin{equation*}
\left[
\begin{array}{c}
\left\vert \downarrow \right\rangle \\
\left\vert \uparrow \right\rangle%
\end{array}%
\right] =\widehat{S}_{1}\left[
\begin{array}{c}
\left\vert 0\right\rangle \\
\left\vert 1\right\rangle%
\end{array}%
\right] =\widehat{S}_{1}\widehat{S}^{-1}\left[
\begin{array}{c}
\left\vert -\right\rangle \\
\left\vert +\right\rangle%
\end{array}%
\right] \equiv \widehat{S}_{2}\left[
\begin{array}{c}
\left\vert -\right\rangle \\
\left\vert +\right\rangle%
\end{array}%
\right] .
\end{equation*}%
Assuming $x_{off}=0$, we obtain
\begin{equation*}
\widehat{S}_{2}=\frac{1}{\sqrt{2}}\left[
\begin{array}{cc}
\cos \frac{\eta }{2}+\sin \frac{\eta }{2} & -\cos \frac{\eta }{2}+\sin \frac{%
\eta }{2} \\
\cos \frac{\eta }{2}-\sin \frac{\eta }{2} & \cos \frac{\eta }{2}+\sin \frac{%
\eta }{2}%
\end{array}%
\right] .
\end{equation*}%
Thus, provided we calculate the density matrix in the stationary basis $\widehat{\rho
}$, we find it in the adiabatic basis $\widehat{\rho }_{adiab}$,
\begin{equation}
\widehat{\rho }_{adiab}=\widehat{S}_{2}^{-1}\widehat{\rho }\widehat{S}_{2}.
\label{rho-adiab}
\end{equation}%
The initial condition for $\widehat{\rho }$ can be obtained from the initial
condition for $\widehat{\rho }_{adiab}$ by inverting the relation (\ref%
{rho-adiab}).

Let us now consider as an illustrative example the one-sweep process,
\begin{align*}
t& \in \left( -\frac{T}{4},\frac{T}{4}\right) ,\text{ }T=\frac{2\pi }{\omega
}, \\
x(t)& =x_{0}\sin \omega t,\text{ }x(t)\in \left( -x_{0},x_{0}\right) , \\
x_{0}& \gg \Delta \gg \omega ,
\end{align*}%
which corresponds to the LZ\ model \cite{foot2}. We choose the initial condition to be
\begin{equation*}
\widehat{\rho }_{adiab}\left( -\frac{T}{4}\right) =\left[
\begin{array}{cc}
1 & 0 \\
0 & 0%
\end{array}%
\right] ,
\end{equation*}%
which means that the system at $t=-T/4$ is in the lower adiabatic state $%
\left\vert \downarrow \right\rangle $.\ We look for the occupation probability
$P_{\uparrow }=\rho _{adiab}^{22}(t)$ of the upper adiabatic level $\left\vert \uparrow
\right\rangle $, which equals to the LZ probability at the end of the sweep
\cite{foot2}, $P_{\uparrow }(T/4)=P_{LZ}$. Thus, for the functions $X,Y,Z$, introduced
in Sec. II, we get the initial condition
\begin{eqnarray*}
X\left( -\frac{T}{4}\right)  &=&\frac{x_{0}}{\sqrt{\Delta ^{2}+x_{0}^{2}}},%
\text{ }Y\left( -\frac{T}{4}\right) =0,\text{ } \\
Z\left( -\frac{T}{4}\right)  &=&\frac{\Delta }{\sqrt{\Delta ^{2}+x_{0}^{2}}}
\end{eqnarray*}%
and find
\begin{eqnarray*}
P_{\uparrow }(t) &=&\frac{1}{2}-\frac{\Delta }{2\sqrt{\Delta ^{2}+x(t)^{2}}}%
Z(t)+ \\
&&+\frac{x(t)}{2\sqrt{\Delta ^{2}+x(t)^{2}}}X(t).
\end{eqnarray*}%
This is illustrated in Fig. \ref{P(t)}b which is equivalent to Fig. 3d in Ref. 10.

In the general case, $x_{off}\neq 0$, we calculate $\widehat{S}_{2}$ and then
$\widehat{\rho }_{adiab}$ according to Eq. (\ref{rho-adiab}). Then the probability
$P_{\uparrow }$ is given by
\begin{eqnarray*}
P_{\uparrow }(t) &=&\frac{1}{2}-\frac{(\Delta ^{2}+x(t)x_{off})}{2\sqrt{%
\Delta ^{2}+x(t)^{2}}\sqrt{\Delta ^{2}+x_{off}^{2}}}Z(t)+ \\
&&+\frac{(\Delta x(t)-\Delta x_{off})}{2\sqrt{\Delta ^{2}+x(t)^{2}}\sqrt{%
\Delta ^{2}+x_{off}^{2}}}X(t).
\end{eqnarray*}%
This should be supplemented with the respective initial condition.

\subsection{Crossover from multiphoton Rabi resonances to LZ-interferometry:
numerical results}

Now making use of the numerical solution of Eqs. (\ref{eq1}--\ref{eq3}) for the
Hamiltonian (\ref{Ham_Rabi}), we study the dependence of the time-averaged probability
$\overline{P}_{+}$ on frequency $\omega $ and amplitude $x_{0}$. For small amplitudes,
$x_{0}\ll \Delta E$, there are resonant peaks in the $\overline{P}_{+}$--$\omega $
dependence at $\omega \simeq \Delta E/K$, as described in Sec. III.A$\frac{{}}{{}}$ and
illustrated in\ Fig. \ref{P(w)}. With increasing amplitude $x_{0}$, the resonances
shift to higher frequencies. For $x_{off}=0$, the resonances appear at "odd"\
frequencies ($K=1,3,5,...$) only, as it was studied in Ref. 9. For $x_{off}\neq 0$
there are also resonances at "even"\ frequencies ($K=2,4,...$), which is demonstrated
in Fig. \ref{P(w)}. We note that Fig. \ref{P(w)} is plotted for the ideal case of the
absence of decoherence and relaxation, $\Gamma _{\phi }=\Gamma _{relax}=0$, when the
resonant value is $\overline{P}_{+}=1/2$. The effect of finite dephasing, $%
\Gamma _{\phi }\neq 0$, and relaxation, $\Gamma _{relax}\neq 0$, is to decrease the
resonant values of $\overline{P}_{+}$ and to widen the peaks for $\Gamma _{\phi
}>\Gamma _{relax}$. Thus from the comparison of the theoretically calculated resonant
peaks with the experimentally observed ones, the dephasing $\Gamma _{\phi }$ and the
relaxation rates $\Gamma _{relax}$\ can be obtained \cite{yaponci}.

When the system is driven slowly, $\omega \ll \Delta $, and with large amplitude,
$x_{0}\gg \Delta $, the LZ excitation mechanism is relevant for the description of the
system dynamics. In the previous subsection we considered the LZ transition for a
single-sweep event. Now we study the periodical driving of the system: Interferences
between multiple LZ
transitions happen, which leads to resonant excitations \cite{TerNak}$^{,}$ \cite%
{ShIF}$^{,}$ \cite{BVK}. We will compare these resonances with the multiphoton Rabi
ones.

First, we note that the resonance positions depend on the amplitude; we
demonstrate that for $\omega \ll \Delta <x_{0}$ in the inset of Fig. \ref%
{P(w)}. But it is more illustrative to study the resonance properties via the
dependence of $\overline{P}_{+}$ on the ratio $x_{0}/\omega $ (or, more
precisely, on the phase the state vector picks up per half-period \cite{ShIF}%
\ $\phi =4x_{0}/\omega $). It is reasonable to carry out the calculation for a fixed
value of the product $x_{0}\omega $, which in its turn defines the LZ-probability (see
Eq. (\ref{P_LZ})). This allows us not only to compare our results with the results of
Ref. 12, where the periodicity of the resonances in $\phi $ was predicted, but also to
demonstrate the transition from the multiphoton Rabi resonances to the ones induced due
to the interference of LZ excitations by means of the $\overline{P}_{+}$--$\phi $
dependence.

\begin{figure}[t]
\includegraphics[width=8.0cm]{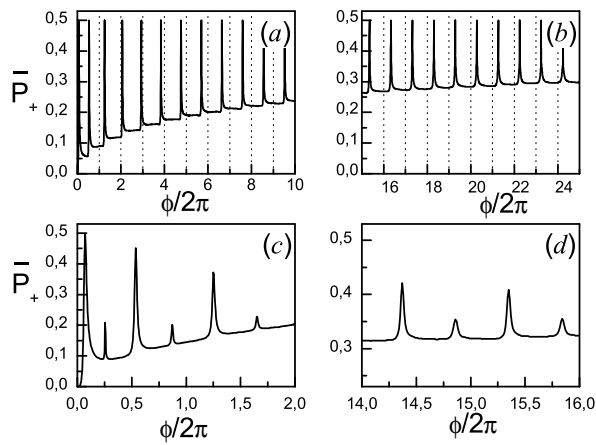}
\caption{Dependence of the probability $\overline{P}_{+}$ on $%
\protect\phi =4x_{0}/\protect\omega $ at $x_{0}\protect\omega /\Delta
^{2}=0.45$ (which corresponds to $P_{LZ}=10^{-3}$); $\Gamma _{\protect\phi %
}=\Gamma _{relax}=0$ and $x_{off}=0$ for graphs (a) and (b); $\Gamma _{%
\protect\phi }/\Delta $ $=\Gamma _{relax}/\Delta =10^{-3}$ and $%
x_{off}/\Delta =0.05$ for graphs (c) and (d).} \label{P(fi)}
\end{figure}

In Fig. \ref{P(fi)}(a,b) we plotted the dependence of $\overline{P}_{+}$ on $%
\phi $ for $x_{off}=0$ and $\Gamma _{\phi }=\Gamma _{relax}=0$. The first few peaks, in
the region $x_{0}\lesssim \Delta $, corresponding to the multiphoton Rabi resonances
are situated at $\phi \sim K^{2}$ (that follows from the resonance relation $\omega
=\Delta E/K$). With increasing $\phi $ (which is proportional to $x_{0}$) we observe
the overall rising of the curve in accordance with the conclusions of Ref. 9. At $\phi
\gg 1$, i.e., at $\omega \ll \Delta \ll x_{0}$, the resonance position is $2\pi $%
-periodic in agreement with Ref. 12. The non-zero offset $x_{off}\neq 0$ results in the
appearance of additional resonances between the basic ones \cite{ShIF}, which is
demonstrated in Fig. \ref{P(fi)}(c,d). Such a feature is similar to the multiphoton
Rabi resonances. Nonzero decoherence decreases and widens the peaks we have also
demonstrated in Fig. \ref{P(fi)}(c,d).

\begin{figure}[t]
\includegraphics[width=3.5cm]{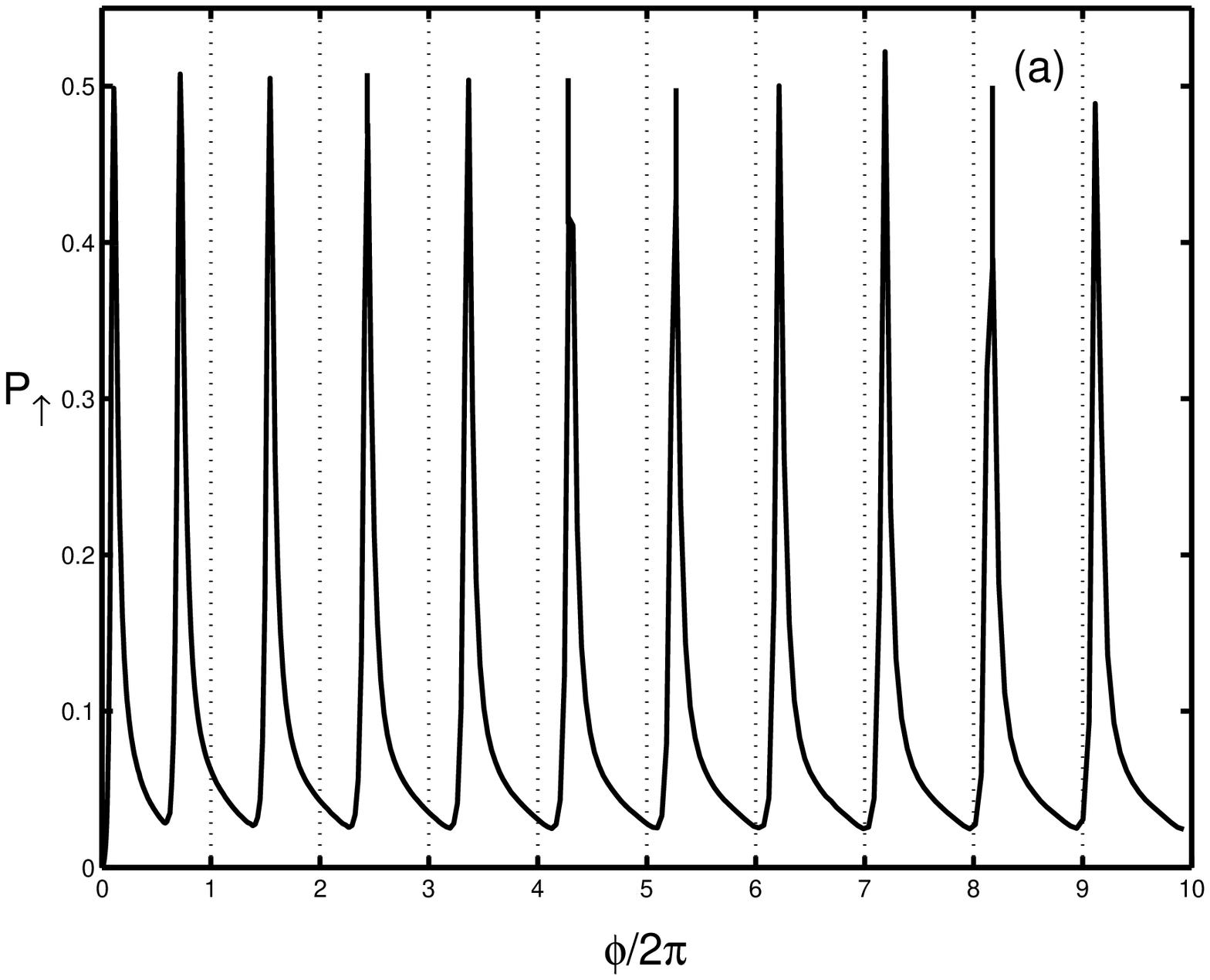}
\includegraphics[width=3.5cm]{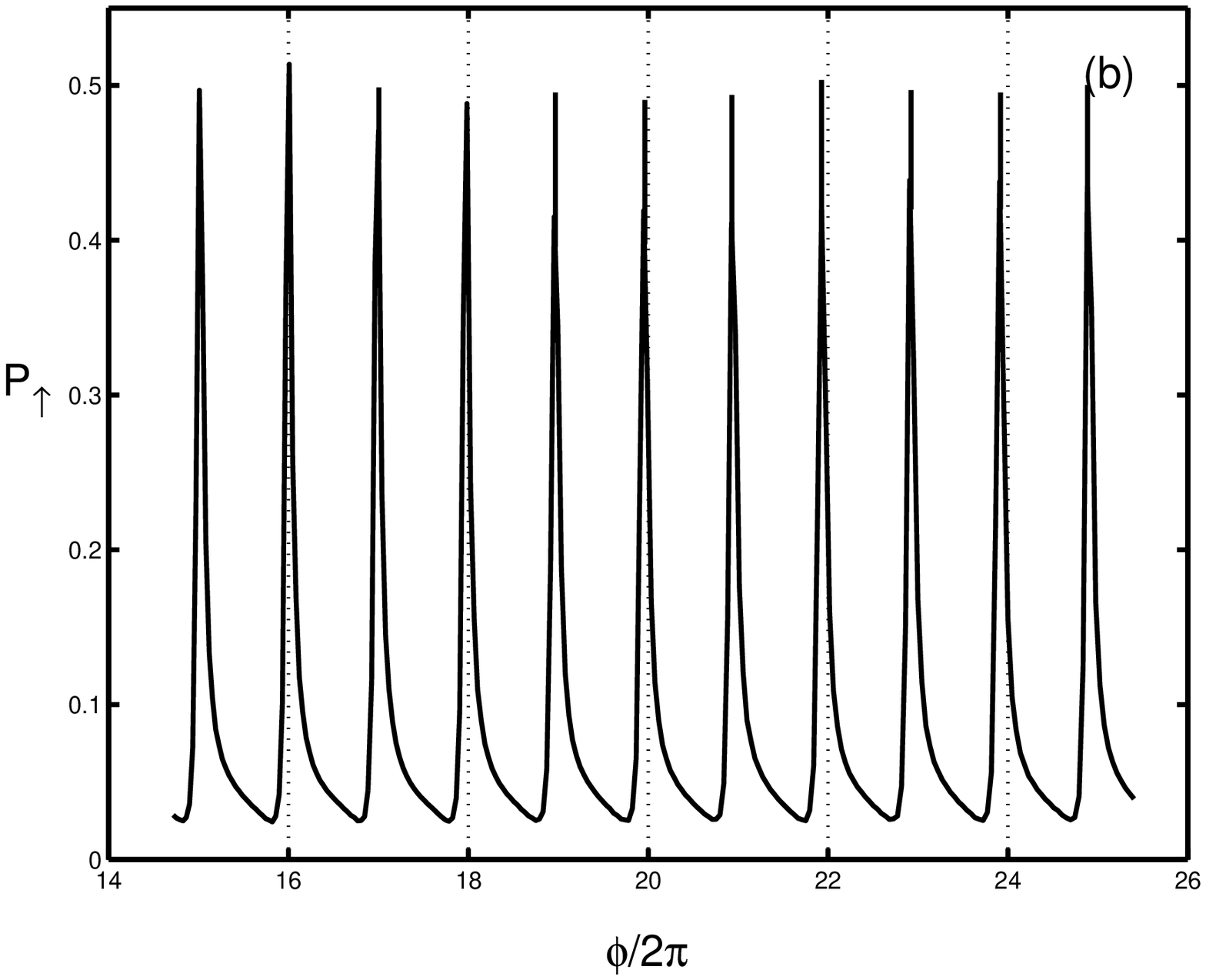}
\includegraphics[width=3.5cm]{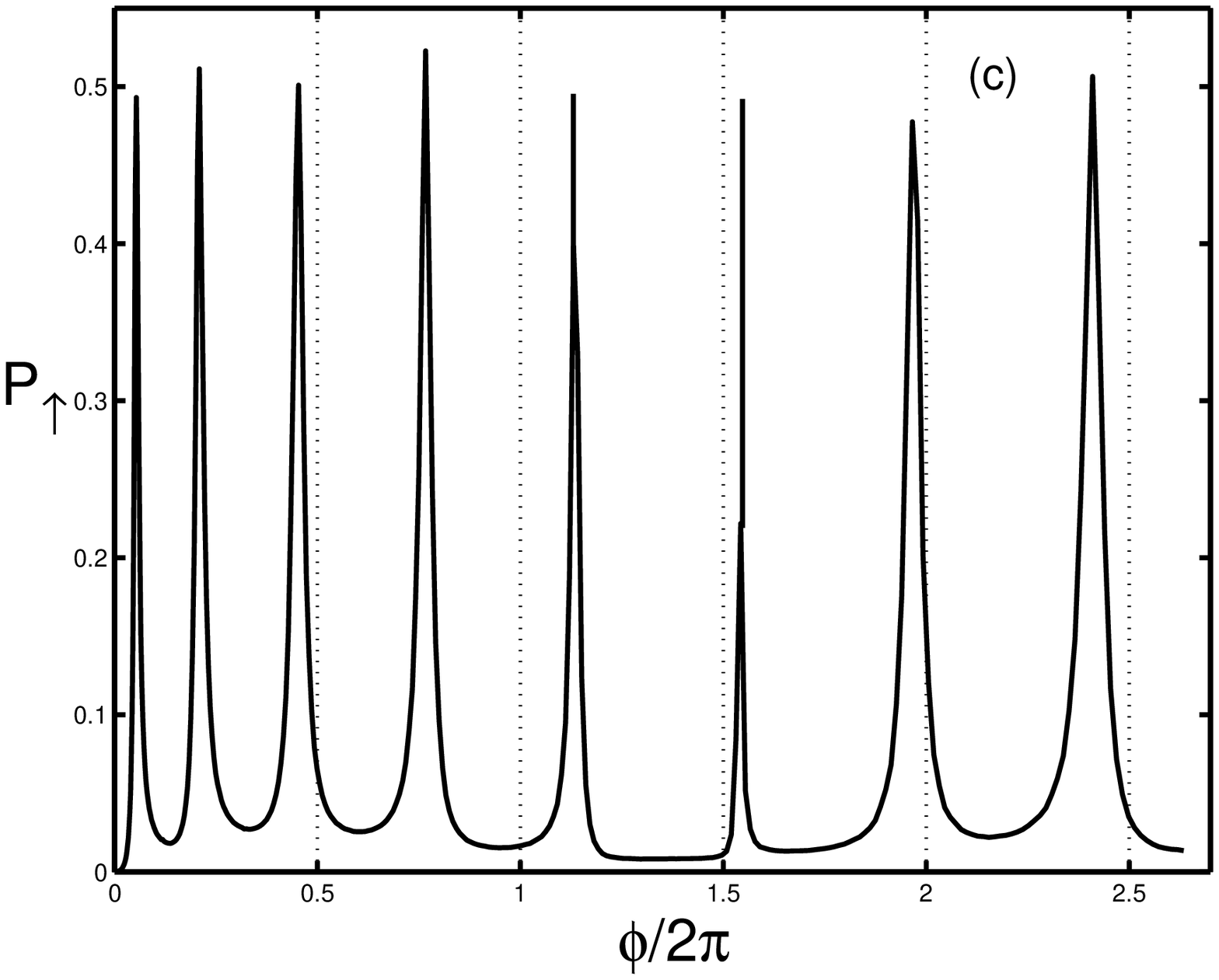}
\includegraphics[width=4cm]{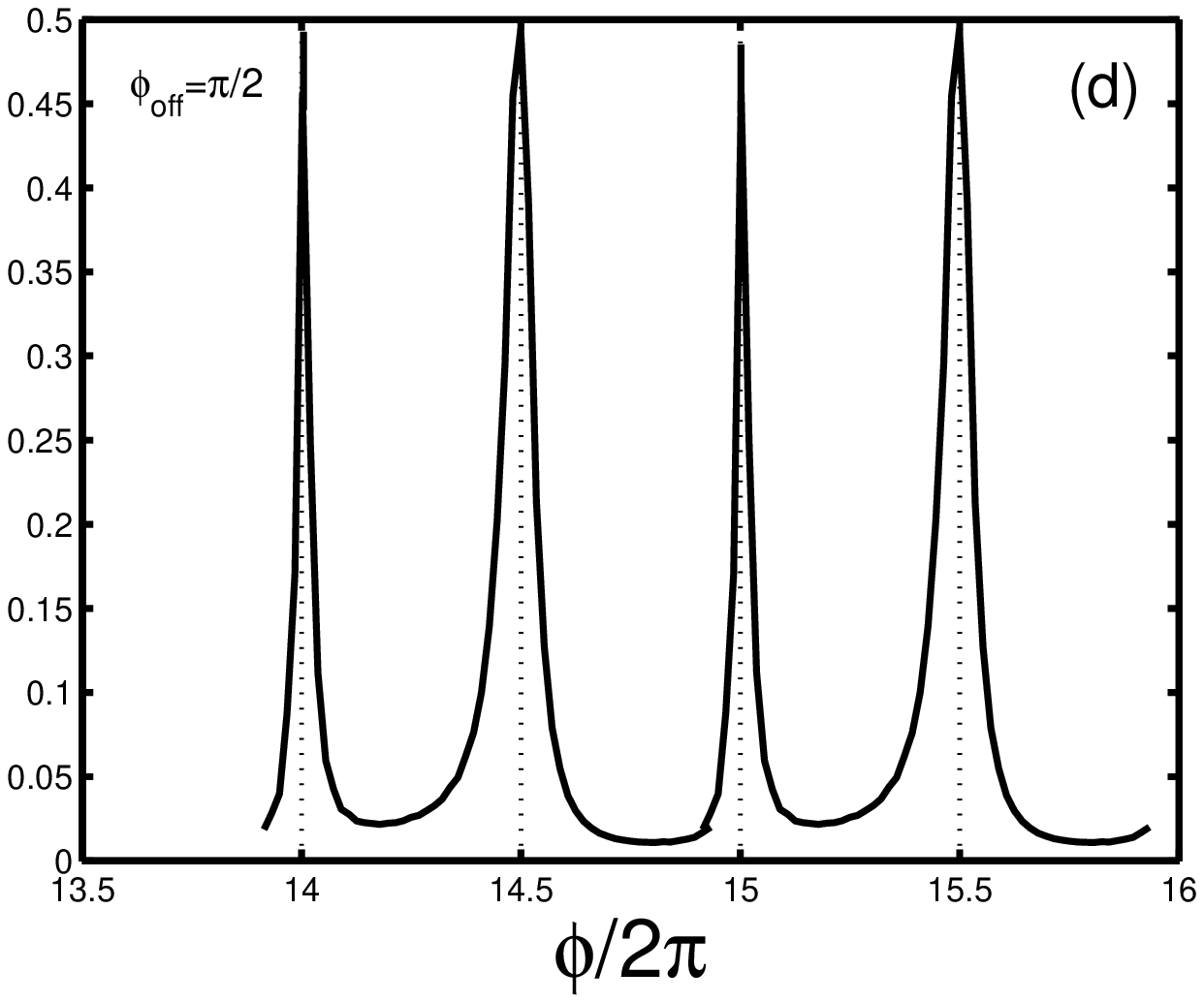}
\caption{The dependence of the probability $%
\overline{P}_{\uparrow }$ on the phase $\protect\phi $.} \label{P_up}
\end{figure}

In Fig. \ref{P_up} we present the dependence of $\overline{P}_{\uparrow }$ on $\phi $
for the calculations carried out in the adiabatic basis (cf. Sec. III.B). In Fig.
\ref{P_up}(a,b) we plotted
such a dependence for different values of $\phi $ at$\ x_{off}=0$ and $%
\Gamma _{\phi }=\Gamma _{relax}=0$. For $\phi \gg 1$ (see Fig. \ref{P_up}b) the
resonances are $2\pi $-periodic, and the peaks are situated at the integer values of
$\phi /2\pi $, as predicted in Ref. 12. We also plotted the dependence of
$\overline{P}_{\uparrow }$ on $\phi $ for the non-zero offset $x_{off}\neq 0$ (see Fig.
\ref{P_up}(c,d)), namely for $\phi _{off}=\pi /2$. For this value the additional peaks
become as high as the basic ones.

\section{Multiphoton excitations in the interferometer-type charge qubit}

\subsection{Interferometer-type charge qubit}

In this section, we consider the quantum dynamics of the interferometer-type
charge qubit with periodically varying control parameters \cite%
{Zorin(02)}$^{-}$\cite{Krech02}. This qubit consists of two Josephson junctions closed
by a superconducting ring. The charge $en$ of the island between the junctions is
controlled by the gate voltage $V_{g}$ via the
capacitance $C_{g}$. The junctions are characterized by Josephson energies $%
E_{J1}$, $E_{J2}$ and phase differences $\varphi _{1}$, $\varphi _{2}$.
The relevant energy values are the island's Coulomb energy, $%
E_{C}=e^{2}/2C_{tot}$, where $C_{tot}$\ is the total capacitance of the island, and the
effective Josephson energy
\begin{equation*}
\varepsilon _{J}=\left( E_{J1}^{2}+E_{J2}^{2}+2E_{J1}E_{J2}\cos \delta \right) ^{1/2}.
\end{equation*}%
An important feature of the qubit is that its Josephson energy is controlled by the
external magnetic flux $\Phi _{e}$ piercing the ring. In this paper, the ring
inductance $L$ is assumed to be small. Then the total phase
difference, $\delta =\varphi _{1}+\varphi _{2}$, approximately equals to $%
\delta _{e}=2\pi \Phi _{e}/\Phi _{0}$, and thus $\varepsilon _{J}=\varepsilon
_{J}(\delta )\simeq \varepsilon _{J}(\delta _{e})$.

Within the two-level model with the basis of "charge\ states" $\left\vert
0\right\rangle $ and $\left\vert 1\right\rangle $ corresponding to the number of Cooper
pairs on the island, the Hamiltonian of the
interferometer-type charge qubit can be written as \cite{MShSh}%
\begin{equation}
\widehat{H}=-\frac{1}{2}\varepsilon _{J}\widehat{\sigma }_{x}-\frac{1}{2}%
E_{ch}\widehat{\sigma }_{z},  \label{Ham_ch_qb_2}
\end{equation}%
where $E_{ch}=4E_{C}(1-n_{g})$ and $n_{g}=C_{g}V_{g}/e$. Here the domination
of the Coulomb energy of a Cooper pair $4E_{C}$ over the coupling energy $%
\varepsilon _{J}$ is assumed, $4E_{C}/\varepsilon _{J}>1$. The eigenstates, $%
\{\left\vert -\right\rangle ,\left\vert +\right\rangle \}$, of this Hamiltonian are
discriminated by the direction of the supercurrent in the ring \cite{Krech02}.

The qubit is considered to be coupled not only to the gate but also to the tank circuit
that enables both phase control and readout. Thus there are two
possibilities to make the Hamiltonian of the two-level system (\ref%
{Ham_ch_qb_2}) time-dependent. First, the gate voltage $V_{g}$ can be driven,
\begin{equation}
n_{g}=n_{g}^{(0)}+n_{g}^{(1)}\sin \omega t,  \label{n}
\end{equation}%
and, second, the DC and AC components of the current in the tank circuit can induce a
periodically varying magnetic flux,
\begin{equation}
\delta =\delta _{DC}+\delta _{AC}\sin \omega t.  \label{delta}
\end{equation}%
Here we restrict our consideration to the case of the sinusoidal time dependence of the
parameters.

Further we study the time-averaged occupation probability of the excited state
$\overline{P}_{+}$. We note that because the two states $\left\vert -\right\rangle $\
and $\left\vert +\right\rangle $\ belong to oppositely circulating currents in the
ring, they correspond to different signs of the qubit's Josephson inductance. The
latter can be probed by the impedance measuring technique \cite{Ilichev04}, which makes
it possible to observe the resonant behaviour of $\overline{P}_{+}$ studied in the
following subsection.

\subsection{Resonant excitation of the interferometer-type charge qubit}

The case of the excitation of the interferometer-type charge qubit via the gate (Eq.
(\ref{n})) can easily be related to the one considered in Sec. III; see the
Hamiltonians (\ref{Ham_ShIF}) and (\ref{Ham_ch_qb_2}) with the
diagonal time-dependent parameters defined by Eq. (\ref{n}). Then the ratio $%
x_{0}/\Delta $ is given by
\begin{equation*}
\frac{x_{0}}{\Delta }=\frac{4E_{C}}{\varepsilon _{J}}n_{g}^{(1)}.
\end{equation*}%
Thus, both mechanisms considered in Sec. III can be realized: multiphoton excitations
($x_{0}/\Delta \ll 1$) and LZ interferometry ($x_{0}/\Delta \,\gg 1$).

Let us now consider the second possibility of the excitation of the qubit by the
time-dependent magnetic flux. The Hamiltonian of the interferometer-type charge qubit
(\ref{Ham_ch_qb_2}) with the periodically varying phase $\delta $ (Eq. (\ref{delta}))
is related to the Hamiltonian of a two-level system considered in Sec. II (Eq.
(\ref{H2})) by the following relations:
\begin{align*}
B_{x}(t)& \equiv \varepsilon _{J}, \\
B_{x}^{(0)}& =B_{x}(0)\equiv \left. B_{x}\right\vert _{\delta _{AC}=0}, \\
B_{x}^{(1)}(t)& =B_{x}(t)-B_{x}(0), \\
B_{z}^{(0)}& =E_{ch}.
\end{align*}

In two limiting cases, the expressions can be simplified to result the time-dependent
term in the form
\begin{equation}
B_{x}^{(1)}(t)\propto \sin \omega t.  \label{b}
\end{equation}

Namely, for
\begin{equation}
\delta -\pi \gg \frac{\left\vert E_{J1}-E_{J2}\right\vert }{\sqrt{%
E_{J1}E_{J2}}},\text{ }\delta _{AC}\lesssim \delta _{DC}-\pi \ll \pi \label{i}
\end{equation}%
we have
\begin{align}
B_{x}^{(0)}& \simeq \sqrt{E_{J1}E_{J2}}\left( \delta _{DC}-\pi \right) ,
\label{i1} \\
B_{x}^{(1)}(t)& \simeq \sqrt{E_{J1}E_{J2}}\delta _{AC}\sin \omega t, \label{i2}
\end{align}%
and for
\begin{equation}
\delta -\pi \ll \frac{\left\vert E_{J1}-E_{J2}\right\vert }{\sqrt{%
E_{J1}E_{J2}}},\text{ }\delta _{AC}\ll \delta _{DC}-\pi \ll \pi  \label{ii}
\end{equation}%
it follows
\begin{align}
B_{x}^{(0)}& \simeq \left\vert E_{J1}-E_{J2}\right\vert ,  \label{ii1} \\
B_{x}^{(1)}(t)& \simeq \frac{E_{J1}E_{J2}}{\left\vert E_{J1}-E_{J2}\right\vert }(\delta
_{DC}-\pi )\delta _{AC}\sin \omega t. \label{ii2}
\end{align}

When the relation (\ref{b}) takes place, the present problem (of the charge qubit with
time-dependent magnetic control) can be related to the problem considered above in Sec.
III by the expressions (\ref{substitute}). Then we can estimate the ratio $x_{0}/\Delta
E$, which characterizes the mechanism of the excitation of the qubit. E.g., at
$x_{0}/\Delta E\lesssim 1$ we expect multiphoton resonances in the qubit's response to
the external alternating magnetic flux. We consider this case below in detail.

We note that the width of the resonant peaks is defined by the non-diagonal components
of the Hamiltonian (\ref{Ham_Rabi}), i.e. by the product $x_{0}\Delta $. Then looking
at the Eqs. (\ref{i2}) and (\ref{ii2}), we conclude that the width of the resonances is
defined by the product $\delta _{AC}(1-n_{g})$.

\begin{figure}[t]
\includegraphics[width=7.0cm]{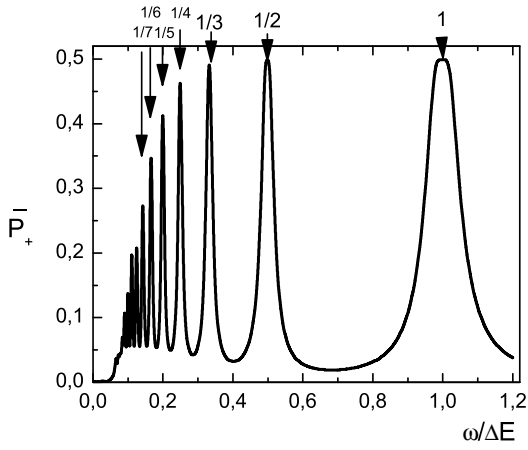}
\caption{Dependence of the probability $\overline{P}_{+}$ on the frequency
$\protect\omega $ for the
phase-biased charge qubit at $n_{g}=0.95$, $E_{J1}/E_{C}=12.4$, $%
E_{J2}/E_{C}=11$, $\Gamma _{\phi }/E_{C}=5\cdot 10^{-4}$, $\Gamma _{relax}/E_{C}%
=10^{-4}$, $\protect\delta _{AC}=0.2\protect\pi $, $\protect\delta _{DC}=%
\protect\pi +0.2\protect\pi $.} \label{P(w)_2}
\end{figure}

Finally, we illustrate the multiphoton resonant excitations of the interferometer-type
charge qubit. Making use of the numerical solution of
the master equation (Sec. II), we find the time-averaged probability $%
\overline{P}_{+}$ plotted in Fig. \ref{P(w)_2}. The position of the multiphoton
resonant peaks is defined by the relation $\omega =\Delta E/K$, where $\Delta E=\Delta
E(\delta _{DC})$ is supposed to be fixed. Alternatively, when the $\delta _{DC}$
component of the phase is changed and the frequency $\omega $ is fixed, a similar graph
can be plotted with
resonances at $\delta _{DC}=\delta _{DC}^{(K)}$ defined by the relation $%
\Delta E(\delta _{DC}^{(K)})=K\omega $.

\section{Conclusions}

We have studied the dynamic behaviour of a quantum two-level system subjected to
periodical sweeping of its parameters. The energy levels population was calculated by
solving the master equation for the density matrix with relaxation terms. Studying the
population of the excited state in both stationary and adiabatic bases, we analyzed
some features of the multiphoton Rabi and LZ effects. Particularly, we have shown
certain similarities of the multiphoton resonances at $x_{0}\ll \omega $ with the
resonances at\ $x_{0}\gg \Delta \gg \omega $ due to the interference between multiple
LZ transitions. Based on the solution of the master equation for the density matrix, we
described in detail the dynamic behaviour of the interferometer-type charge qubit
subjected to periodically changing gate voltage or magnetic flux.
%
%

This work was supported in part by the program \textquotedblright Nanosystems,
nanomaterials, and nanotechnology\textquotedblright\ of the National Academy of
Sciences of Ukraine. S.N. Sh. acknowledges the support of the German Academic Exchange
Service (DAAD). The authors thank V.I. Shnyrkov for fruitful discussions.


\begin{thebibliography}{99}
\bibitem{LZS} L.D. Landau, Phys. Z. Sowjetunion \textbf{2}, 46 (1932); C.
Zener, Proc. R. Soc. London, Ser. A \textbf{137}, 696 (1932).

\bibitem{BVK} V.A. Benderskii, E.V. Vetoshkin, and E.I. Kats, JETP \textbf{97%
}, 232 (2003).

\bibitem{Rabi} I.I. Rabi, Phys. Rev. \textbf{51}, 652 (1937).

\bibitem{DelKrai} N.B. Delone and V.P. Krainov, \textit{Atoms in Strong
Light Fields}, Springer Ser. Chem. Phys., Vol. 28, Springer, Berlin--Heidelberg (1985)
(also in Russian: \textit{Atom v sil'nom svetovom pole}, Atomizdat, Moscow (1978)).

\bibitem{GrifHang} M. Grifoni and P. H\"{a}nggi, Phys. Rep. \textbf{304},
229 (1998).

\bibitem{MShSh} Yu. Makhlin, G. Sch\"{o}n and A. Shnirman, Rev. Mod. Phys.
\textbf{73}, 357 (2001).

\bibitem{chargeQbit} Y. Nakamura, Yu.A. Pashkin, and J.S. Tsai, Nature
\textbf{398}, 786 (1999).

\bibitem{fluxQbit} J.E. Mooij \textit{et al}., Science \textbf{285}, 1036
(1999).

\bibitem{KrainovYakovlev} V.P. Krainov and V.P. Yakovlev, Sov. Zh. Exp.
Theor. Fiz. \textbf{78}, 2204 (1980).

\bibitem{Vitanov} N.V. Vitanov, Phys. Rev. A \textbf{59}, 988 (1999).

\bibitem{GaraninShilling} D.A. Garanin and R. Schilling, Phys. Rev. B
\textbf{66}, 174438 (2002).

\bibitem{ShIF} A.V. Shytov, D.A. Ivanov, and M.V. Feigel'man,
cond-mat/0110490 v2.

\bibitem{TerNak} Y. Teranishi and H. Nakamura, Phys. Rev. Lett. \textbf{81},
2032 (1998).

\bibitem{Nakamura01} Y. Nakamura, Yu.A. Pashkin, and J.S. Tsai, Phys. Rev.
Lett. \textbf{87}, 246601 (2001).

\bibitem{You} J.Q. You and F. Nori, Phys. Rev. B \textbf{68}, 064509 (2003).

\bibitem{yaponci} S. Saito \textit{et al}., Phys. Rev. Lett. \textbf{93},
037001 (2004).

\bibitem{Born} D. Born \textit{et al}., Phys. Rev. B \textbf{70}, 180501(R)
(2004).

\bibitem{Falci} G. Falci \textit{et al}., Nature \textbf{407}, 355 (2000).

\bibitem{Zorin(02)} A.B. Zorin, Physica C \textbf{368}, 284 (2002).

\bibitem{FriedmanAverin} J.R. Friedman and D.V. Averin, Phys. Rev. Lett.
\textbf{88}, 050403 (2002).

\bibitem{Krech02} W. Krech \textit{et al}., Phys. Lett. A \textbf{303}, 352
(2002).

\bibitem{Blum} K. Blum, \textit{Density Matrix Theory and Applications},
Plenum Press, New York--London (1981).

\bibitem{foot1} The phenomenological description of the relaxation processes
in terms of the rates was shown \cite{ShIF} to be equivalent to the microscopic
description which intrinsically includes the reservoir. Such a phenomenological
approach was successfully applied to the description of the quantum dynamics of phase
qubits \cite{yaponci}.

\bibitem{Aasime} A. Aassime \textit{et al.}, Phys. Rev. Lett. \textbf{86},
3376 (2001).

\bibitem{Ilichev04} E. Il'ichev \textit{et al}., Low Temp. Phys. \textbf{30}
, 620 (2004).

\bibitem{Flugge} S. Fl\"{u}gge, \textit{Practical Quantum Mechanics
} II, Springer, Berlin--Heidelberg--New York (1971), task 180.

\bibitem{CohTan} C. Cohen-Tannoudji, J. Dupont-Roc, and G. Grynberg, \textit{
Atom-Photon Interactions}, Wiley, New York (1992).

\bibitem{foot2} In the vicinity of the transition point $x=0$ (where the
excitation happens), $x(t)$ is a linear function of $t$, $x(t)\simeq
x_{0}\omega t$, which is exactly the LZ model. Deviations of the function $%
x(t)$ from the linear dependence far from the transition point do not matter
\cite{GaraninShilling}. This explains the quantitative coincidence of the results
presented in this section with the results obtained with $x(t)$ being a linear
function.
\end{thebibliography}
\end{document}